\documentclass[copyright,creativecommons]{eptcsstyle/eptcs}
\providecommand{\event}{QPL 2017} 
\usepackage{xcolor}
\hypersetup{
    colorlinks,
    linkcolor={red!50!black},
    citecolor={blue!50!black},
    urlcolor={blue!80!black}
}
\usepackage[capitalize,noabbrev]{cleveref} 

\usepackage[utf8]{inputenc}
\newcommand{\upsidedownbang}{\rotatebox[origin=c]{180}{!}}
\usepackage{anyfontsize}
\usepackage{xspace}

\usepackage{wrapfig}

\usepackage[inline,shortlabels]{enumitem}

\usepackage{mathpartir}

\usepackage{cmll} 
\usepackage{ stmaryrd } 
\usepackage{upquote} 

\usepackage[numbers]{natbib}

\usepackage[qm,braket]{qcircuit}
\newcommand{\qwire}{\ensuremath{\mathcal{Q}\textsc{wire}}\xspace}
\newcommand{\liquid}{LIQ\emph{Ui}$\ket{}$}

\usepackage{tikz} 
\usetikzlibrary{%
  arrows,%
  shapes.misc,
  shapes.arrows,%
  shapes.callouts,
  chains,%
  matrix,%
  positioning,
  scopes,%
  decorations.pathmorphing,
  decorations.text,
  shadows%
}
\tikzset{ machine/.style={
    rectangle,
    minimum width=20mm,
    minimum height=10mm,
    text width=16mm,
    align=center,
    very thick,
    draw=black,
    color=black,
    fill=white,
    font=\ttfamily,
  }
}

\newcommand*\rfrac[2]{{}^{#1}\!/_{#2}}

\newcommand{\etal}{\emph{et al.}\xspace}
\newcommand{\eg}{\emph{e.g.,}\xspace}

\newcommand{\lbind}{\ensuremath{\mathbin{=\!\!>\!\!>\!\!=}}}

\usepackage{booktabs}

\usepackage{listings,lstcoq}
\usepackage{MnSymbol}
\definecolor{ltblue}{rgb}{0,0.4,0.4}
\definecolor{dkblue}{rgb}{0,0.1,0.6}
\definecolor{dkgreen}{rgb}{0,0.35,0}
\definecolor{dkviolet}{rgb}{0.3,0,0.5}
\definecolor{dkred}{rgb}{0.5,0,0}
\newcommand{\code}[1]{{\small\texttt{#1}}}

\newenvironment{nscenter}
 {\parskip=3pt\par\nopagebreak\centering}
 {\par\noindent\ignorespacesafterend}

\usepackage{newunicodechar}
\let\Alpha=A
\let\Beta=B
\let\Epsilon=E
\let\Zeta=Z
\let\Eta=H
\let\Iota=I
\let\Kappa=K
\let\Mu=M
\let\Nu=N
\let\Omicron=O
\let\omicron=o
\let\Rho=P
\let\Tau=T
\let\Chi=X

\newunicodechar{Α}{\ensuremath{\Alpha}}
\newunicodechar{α}{\ensuremath{\alpha}}
\newunicodechar{Β}{\ensuremath{\Beta}}
\newunicodechar{β}{\ensuremath{\beta}}
\newunicodechar{Γ}{\ensuremath{\Gamma}}
\newunicodechar{γ}{\ensuremath{\gamma}}
\newunicodechar{Δ}{\ensuremath{\Delta}}
\newunicodechar{δ}{\ensuremath{\delta}}
\newunicodechar{Ε}{\ensuremath{\Epsilon}}
\newunicodechar{ε}{\ensuremath{\epsilon}}
\newunicodechar{ϵ}{\ensuremath{\varepsilon}}
\newunicodechar{Ζ}{\ensuremath{\Zeta}}
\newunicodechar{ζ}{\ensuremath{\zeta}}
\newunicodechar{Η}{\ensuremath{\Eta}}
\newunicodechar{η}{\ensuremath{\eta}}
\newunicodechar{Θ}{\ensuremath{\Theta}}
\newunicodechar{θ}{\ensuremath{\theta}}
\newunicodechar{ϑ}{\ensuremath{\vartheta}}
\newunicodechar{Ι}{\ensuremath{\Iota}}
\newunicodechar{ι}{\ensuremath{\iota}}
\newunicodechar{Κ}{\ensuremath{\Kappa}}
\newunicodechar{κ}{\ensuremath{\kappa}}
\newunicodechar{Λ}{\ensuremath{\Lambda}}
\newunicodechar{λ}{\ensuremath{\lambda}}
\newunicodechar{Μ}{\ensuremath{\Mu}}
\newunicodechar{μ}{\ensuremath{\mu}}
\newunicodechar{Ν}{\ensuremath{\Nu}}
\newunicodechar{ν}{\ensuremath{\nu}}
\newunicodechar{Ξ}{\ensuremath{\Xi}}
\newunicodechar{ξ}{\ensuremath{\xi}}
\newunicodechar{Ο}{\ensuremath{\Omicron}}
\newunicodechar{ο}{\ensuremath{\omicron}}
\newunicodechar{Π}{\ensuremath{\Pi}}
\newunicodechar{π}{\ensuremath{\pi}}
\newunicodechar{ϖ}{\ensuremath{\varpi}}
\newunicodechar{Ρ}{\ensuremath{\Rho}}
\newunicodechar{ρ}{\ensuremath{\rho}}
\newunicodechar{ϱ}{\ensuremath{\varrho}}
\newunicodechar{Σ}{\ensuremath{\Sigma}}
\newunicodechar{σ}{\ensuremath{\sigma}}
\newunicodechar{ς}{\ensuremath{\varsigma}}
\newunicodechar{Τ}{\ensuremath{\Tau}}
\newunicodechar{τ}{\ensuremath{\tau}}
\newunicodechar{Υ}{\ensuremath{\Upsilon}}
\newunicodechar{υ}{\ensuremath{\upsilon}}
\newunicodechar{Φ}{\ensuremath{\Phi}}
\newunicodechar{φ}{\ensuremath{\phi}}
\newunicodechar{ϕ}{\ensuremath{\varphi}}
\newunicodechar{Χ}{\ensuremath{\Chi}}
\newunicodechar{χ}{\ensuremath{\chi}}
\newunicodechar{Ψ}{\ensuremath{\Psi}}
\newunicodechar{ψ}{\ensuremath{\psi}}
\newunicodechar{Ω}{\ensuremath{\Omega}}
\newunicodechar{ω}{\ensuremath{\omega}}

\newunicodechar{ℕ}{\ensuremath{\mathbb{N}}}
\newunicodechar{∅}{\ensuremath{\emptyset}}

\newunicodechar{•}{\ensuremath{\bullet}}
\newunicodechar{≈}{\ensuremath{\approx}}
\newunicodechar{≅}{\ensuremath{\cong}}
\newunicodechar{≡}{\ensuremath{\equiv}}
\newunicodechar{≤}{\ensuremath{\le}}
\newunicodechar{≥}{\ensuremath{\ge}}
\newunicodechar{≠}{\ensuremath{\neq}}
\newunicodechar{∀}{\ensuremath{\forall}}
\newunicodechar{∃}{\ensuremath{\exists}}
\newunicodechar{±}{\ensuremath{\pm}}
\newunicodechar{∓}{\ensuremath{\pm}}
\newunicodechar{·}{\ensuremath{\cdot}}
\newunicodechar{⋯}{\ensuremath{\cdots}}
\newunicodechar{…}{\ensuremath{\ldots}}
\newunicodechar{∷}{~\mathrel{:\!\!\!:}~}
\newunicodechar{×}{\ensuremath{\times}}
\newunicodechar{∞}{\ensuremath{\infty}}
\newunicodechar{→}{\ensuremath{\to}}
\newunicodechar{←}{\ensuremath{\leftarrow}}
\newunicodechar{⇒}{\ensuremath{\Rightarrow}}
\newunicodechar{↦}{\ensuremath{\mapsto}}
\newunicodechar{↝}{\ensuremath{\leadsto}}
\newunicodechar{∨}{\ensuremath{\vee}}
\newunicodechar{∧}{\ensuremath{\wedge}}
\newunicodechar{⊢}{\ensuremath{\vdash}}
\newunicodechar{⊣}{\ensuremath{\dashv}}
\newunicodechar{∣}{\ensuremath{\mid}}
\newunicodechar{∈}{\ensuremath{\in}}
\newunicodechar{⊆}{\ensuremath{\subseteq}}
\newunicodechar{⋓}{\ensuremath{\Cup}}
\newunicodechar{∉}{\ensuremath{\not\in}}

\newunicodechar{⊸}{\ensuremath{\multimap}}
\newunicodechar{⊗}{\ensuremath{\otimes}}
\newunicodechar{⊕}{\ensuremath{\oplus}}
\newunicodechar{〈}{\ensuremath{\langle}}
\newunicodechar{⟨}{\ensuremath{\langle}}
\newunicodechar{⟩}{\ensuremath{\rangle}}
\newunicodechar{〉}{\ensuremath{\rangle}}
\newunicodechar{¡}{\ensuremath{\upsidedownbang}}
\newunicodechar{∘}{\ensuremath{\circ}}
\newunicodechar{†}{\ensuremath{\dagger}}
\newunicodechar{⊤}{\ensuremath{\top}}

\newunicodechar{〚}{\ensuremath{\llbracket}}
\newunicodechar{〛}{\ensuremath{\rrbracket}}

\usepackage{etoolbox} 
\newtoggle{comments}
\toggletrue{comments}

\iftoggle{comments}{
  \newcommand{\jennifer}[1]{\textbf{\textcolor{red}{[ #1 --- Jennifer]}}}
  \newcommand{\steve}[1]{\textbf{\textcolor{green}{[ #1 --- Steve]}}}
  \newcommand{\robert}[1]{\textbf{\textcolor{blue}{[ #1 --- Robert]}}}

  \newcommand{\fixme}[1]{\textbf{\textcolor{red}{[ Fixme: #1]}}}
  \newcommand{\note}[1]{\textbf{\textcolor{green}{[ Note: #1 ]}}}
  \newcommand{\todo}[1]{\textbf{\textcolor{green}{[ TODO: #1 ]}}}
}{
  \newcommand{\jennifer}[1]{}
  \newcommand{\steve}[1]{}
  \newcommand{\robert}[1]{}

  \newcommand{\fixme}[1]{}
  \newcommand{\note}[1]{}
  \newcommand{\todo}[1]{}
}

\title{\qwire Practice: \\
Formal Verification of Quantum Circuits in Coq
\thanks{This work is supported in part by ONR MURI No. FA9550-16-1-0082
and NSF Grant No. CCF- 1421193.}}
\def\titlerunning{\qwire Practice}

\author{Robert Rand \email{rrand@seas.upenn.edu} \and Jennifer Paykin \email{jpaykin@seas.upenn.edu} \institute{~\\ \large University of Pennsylvania} \and Steve Zdancewic \email{stevez@cis.upenn.edu}}
\def\authorrunning{Rand, Paykin \& Zdancewic}
\renewcommand\footnotemark{} 

\begin{document}

\maketitle


\begin{abstract}

We describe an embedding of the \qwire quantum circuit language in the Coq proof assistant. This allows programmers to write quantum circuits using high-level abstractions and to prove properties of those circuits using Coq’s theorem proving features. The implementation uses higher-order abstract syntax to represent variable binding and provides a type-checking algorithm for linear wire types, ensuring that quantum circuits are well-formed. We formalize a denotational semantics that interprets \qwire circuits as superoperators on density matrices, and prove the correctness of some simple quantum programs.
 

\end{abstract}

\section{Introduction}
\label{sec:introduction}

The last few years have witnessed the emergence of
lightweight, scalable, and expressive quantum circuit languages such as
Quipper~\cite{Green2013} and \liquid~\cite{Wecker2014}. 
These languages adopt the QRAM model of quantum computation, in which a 
classical computer sends instructions to a quantum computer and receives back 
measurement results. 
Quipper and \liquid{} programs
classically produce circuits that can be executed on a quantum computer,
simulated on a classical computer, or compiled using classical techniques to
smaller, faster circuits. Since both languages are embedded
inside general-purpose classical host languages (Haskell and F\#),
they can be used to build useful abstractions on top of quantum
circuits, 
allowing for general purpose quantum programming.


As is the case with classical programs, however, quantum programs in these
languages will invariably have bugs. Since quantum circuits are inherently
expensive to run (either simulated or on a real quantum computer) and
are difficult or impossible to debug at runtime, numerous techniques have been developed to
verify properties of quantum programs. 

The first step towards guaranteeing bug-free quantum programs is ensuring that 
every program corresponds to a valid
quantum computation, meaning that a simulator or quantum computer running that
program will not crash. In many cases this property can be enforced using type
systems, as in the quantum lambda calculus, which uses linear types and
guarantees type safety~\cite{Selinger2009}. Along these lines,
Proto-Quipper~\cite{Ross2015} adds linear types to a subset of Quipper, though
this approach has not been extended to the full Quipper language.

Beyond simply ensuring quantum mechanical soundness,
we might wish to statically analyze specific programs or families of
programs and prove that their semantics matches a formal specification. Doing so
requires a formal semantics for programs, such as unit vectors or density
matrices. 
The specification can then be verified using model-checking~\cite{Gay2008}, Hoare logic~\cite{Ying2011},
proof assistants~\cite{Boender2015}, or other techniques.

Finally, one may wish to verify that two particular programs (or program
fragments) have the same semantics. \liquid, in particular, has focused on
efficient compilation of quantum circuits; similar projects have explored
verified compilation passes in the case of reversible circuits~\cite{Amy2017}.

The \qwire programming language~\cite{Paykin2017} is a small quantum circuit
language embedded in a classical host language, which provides three core features:
\begin{enumerate*}[(1)]
\item a platform for high level quantum computing, with the
expressiveness of embedded languages like Quipper \cite{Green2013} and \liquid
\cite{Wecker2014};
\item a linear type system that guarantees that generated
circuits are well-formed and respect the laws of quantum mechanics; and
\item a concrete denotational semantics, specified in terms of
density matrices, for proving properties and equivalences of quantum circuits.
\end{enumerate*}

In this paper we report on a ongoing effort to implement \qwire in the Coq
theorem prover~\cite{Coq} and formalize its denotational semantics, thereby
providing a framework to formally verify the correctness of quantum programs and
quantum program transformations.\footnote{The Coq development for this paper is
available at: \url{https://github.com/jpaykin/QWIRE/tree/QPL2017}.} We include two 
forms of the language, one using Coq variables and higher-order abstract syntax for
ease of programming, and a simpler version without these features, for verification
purposes, along with a simple translation from the former to the latter. We then use
this bridge to prove properties about higher-order programs using \qwire's denotational
semantics.

The paper makes the following contributions:
\begin{itemize}

\item We implement the \qwire programming language in the Coq proof assistant,
  incorporating features such as dependently-typed circuits and proof-carrying
  code;

\item We present a type checking algorithm for the linear type system of \qwire
  using a representation of linear contexts based on the Linearity
  Monad~\cite{Paykin2017a}; 

\item We formalize a denotational semantics for \qwire circuits, interpreting them
  as superoperators on density matrices~\cite{Nielsen2010}; and

\item We show how these semantics can be used to verify quantum programs,
  including a quantum coin flip, a protocol with \emph{dynamic lifting},
  and a simple unitary circuit. 
\end{itemize}

Throughout the paper, we will assume some level of familiarity with both
functional programming and proof assistants in the style of Coq or Agda. Readers
unfamiliar with either of these concepts should consult the the introductory
chapter of \emph{Software Foundations}~\cite{Pierce2016}, which introduces the
Coq language and theorem prover~\cite{Coq} used here.

\section{Introduction to \qwire programming}
\label{sec:qwire-examples}

We start with some examples of \qwire circuits implemented in Coq. The following
circuit implements a quantum coin flip:\footnote{Unless otherwise indicated, all
  \qwire circuit definitions end in an implicit \coqe{Defined}.}

\begin{minipage}{0.6\textwidth}
\begin{coq}
Definition coin_flip : Box One Bit.
  box_ () =>
    gate_ x  <- init0 @();
    gate_ y  <- H @x;
    gate_ z  <- meas @y;
    output z.
\end{coq}
\end{minipage}
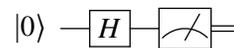
\begin{minipage}{0.4\textwidth} \begin{center}
\Qcircuit @C=1em @R=.7em {
    \lstick{\ket{0}} & \gate{H} & \meter & \cw
}
\end{center} \end{minipage}

The type \code{Box One Bit} is the high-level interface to \qwire
circuits, representing a ``boxed'' circuit with no input wires
(represented by the unit type \coqe{One}) and a \coqe{Bit}-valued
output wire. We use Coq notations and tactics to construct the
circuit. The \coqe{box_} tactic creates a boxed circuit from a
function that takes an input pattern of wires and produces a circuit
that uses those wires; \coqe{box_} also calls the linear type-checker
on the result, to ensure that the circuit is well formed.  In this
example, the \coqe{coin_flip} circuit has no inputs, so the input
pattern to the box is the empty pattern \coqe{()}.

The body of the circuit is made up of a sequence of gate applications, where
\coqe{init0}, \coqe{H}, and \coqe{meas} are all gates. These gates are applied
to input patterns, whose types are determined by the gate being applied. In
\coqe{coin_flip}, the pattern \coqe{()} has type \coqe{One}, \coqe{x} and
\coqe{y} have type \coqe{Qubit}, and \coqe{z} has type \coqe{Bit}. The output of
each gate is bound in the remainder of the circuit, and the circuit is
terminated by an \coqe{output} statement.

\qwire ensures that wires in a circuit are treated linearly, meaning that
every wire is used exactly once as input and once as output. 
For example, the linear type system enforces the no-cloning property, rejecting
the following bogus circuit.
\begin{coq}
Definition clone W : Box W (W ⊗ W). 
  box_ w => output (w,w). (* Linear type checker cannot be satisfied. *) Abort.
\end{coq}

\qwire is compositional, meaning that a boxed circuit can be reused inside
another circuit. Notice that only wires are treated linearly in \qwire; a boxed
circuit is a first-class Coq expression, and can be used as many times as
necessary. Consider the following circuit, which flips a coin up to $n$ times,
returning 1 if and only if the coin always lands on heads, which occurs with probability $\frac{1}{2^n}$.

\hspace*{-\parindent}%
\begin{minipage}{0.68\textwidth}
\begin{coq}
Fixpoint coin_flips (n : nat) : Box One Bit.
  box_ () => match n with    (* n:nat is either 0 or the successor of some n' *)
             | 0    => gate_ x <- new1 @(); output x
             | S n' => let_  c     <- unbox (coin_flips n') ();
                       gate_ q     <- init0 @();
                       gate_ (c,q) <- bit_ctrl H @(c,q);
                       gate_ ()    <- discard @c;
                       gate_ b     <- meas @q;
                       output b
             end.
\end{coq}
\end{minipage}
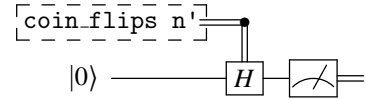
\begin{minipage}{0.2\textwidth}
\small
\vspace{10mm}
\Qcircuit @C=1em @R=1em {
    & \push{\texttt{coin\_flips n\textquotesingle}} & \control \cwx[1] \cw \\
    & \lstick{\ket{0}} & \gate{H} & \meter & \cw \gategroup{1}{2}{1}{2}{.5em}{--}
} \normalsize
\end{minipage}

The \coqe{unbox} operator feeds the input pattern \coqe{()} into the boxed
circuit \coqe{coin_flips n'}, and the \coqe{let_} operation binds the output of
the recursive call in the rest of the circuit. The \coqe{let_} and \coqe{unbox}
operations are separate constructs; for example, \coqe{box_ w ⇒ unbox b w} is
just the $\eta$-expansion of a boxed circuit \coqe{b}.

\qwire also allows for \emph{dynamic lifting}~\cite{Paykin2017}, in which
measurement results from the quantum circuit are dynamically processed as
classical data inside of Coq. This \coqe{lift} operation
measures a qubit (or any collection of wires) and produces a boolean (or a tuple
of booleans). We use the result of the measurement to decide
which gates to apply in the remainder of the circuit. The following variation
on \coqe{coin_flips} uses dynamic lifting and calls the unary \coqe{coin_flip} circuit 
from the start of this section.
\begin{coq}
Fixpoint coin_flips' (n : nat) : Box One Bit.
  box_ () ⇒ match n with
             | 0    => gate_ q ← new1 @(); output q
             | S n' => let_  q ← unbox (coin_flips' n') ();
                       lift_ x ← q;
                       if x then unbox coin_flip () 
                            else gate_ q ← new0 @(); output q
              end. 
\end{coq}

\noindent Further examples can be found in the online development, as well as in the
original \qwire paper~\cite{Paykin2017}.

\section{Implementing \qwire in Coq}
\label{sec:qwire}

At its core, a \qwire circuit is a sequence of gates applied to wires. Each
wire is described by a \emph{wire type} \coqe{W}, which is either the unit type
(has no data), a bit or qubit, or a tuple of wire types. In Coq we represent
wire types as an inductively defined data type \coqe{WType} as follows:
\begin{coq}
Inductive WType := One | Bit | Qubit | Tensor : WType -> WType -> WType.
\end{coq}
We use the Coq notation \coqe{``W1 ⊗ W2''} for \coqe{Tensor W1 W2}.

Gates are indexed by a pair of wire types---a gate of type \coqe{Gate W1 W2}
takes an input wire of type \coqe{W1} and outputs a wire of type \coqe{W2}. In our
setting, gates will include a universal set of unitary gates, as well as
initialization, measurement, and control.\footnote{The set of gates need not be
  fixed; Rennela and Staton~\cite{Rennela2017} explore how to use gates to
  extend \qwire with recursive types.}
\begin{coq}
Inductive Unitary : WType -> Set := (* other unitary gates omitted *)
  | H         : Unitary Qubit (* Hadamard gate *) 
  | control   : forall {W} (U : Unitary W), Unitary (Qubit ⊗ W) 
  | transpose : forall {W} (U : Unitary W), Unitary W.
Inductive Gate : WType -> WType -> Set := 
  | U    : forall {W} (u : Unitary W), Gate W W
  | init : Gate One Qubit 
  | meas : Gate Qubit Bit 
  | discard : Gate Bit One.
\end{coq}
The curly braces surrounding the type argument \coqe{W} indicate that \coqe{W}
is an implicit argument, meaning that it can be automatically inferred from the
other arguments. We further define \coqe{U} to be a \emph{coercion} from
unitaries to gates, meaning that for \coqe{u : Unitary W}, we can simply write
\coqe{u} for \coqe{U u : Gate W W}.

An open  circuit (one with free input wires) has Coq type \coqe{Circuit Γ W},
where $Γ$ is a typing context of input wires and
$W$ is an output wire type. As an example, the circuit
\begin{nscenter}
    \coqe{gate_ w1' ← H @w1; gate_ w2' ← meas @w2; output (w2',w1')}
\end{nscenter}
has type \coqe{Circuit (w1:Qubit, w2:Qubit) (Bit ⊗ Qubit)}.

A typing context of type \coqe{Ctx} is a partial map from variables (represented
concretely as natural numbers) to wire types, which is represented as
\coqe{list (option WType)}.
In this representation, the variable $i$ is mapped to $W$ if the $i$th element
in the list is $\code{Some}~W$, and is undefined if the
$i$th element is $\code{None}$.

The \textit{disjoint merge} operation $\Cup$ ensures that the same wire cannot
be used in two separate parts of a circuit. Mathematically, it is defined on two
typing contexts as follows:
\begin{coq}
[]  ⋓  Γ2                     = Γ2
Γ1  ⋓  []                     = Γ1
None   :: Γ1  ⋓  None   :: Γ2 = None   :: (Γ1 ⋓ Γ2)
Some W :: Γ1  ⋓  None   :: Γ2 = Some W :: (Γ1 ⋓ Γ2)
None   :: Γ1  ⋓  Some W :: Γ2 = Some W :: (Γ1 ⋓ Γ2)
\end{coq}

\noindent Since disjoint merge is a partial function we represent it in Coq as
a relation on possibly invalid contexts, \coqe{OCtx = Invalid | Valid Ctx}.
For convenience, most operations on contexts are lifted to work with
\coqe{OCtx} values, and so the type signature of the merge operation
$⋓$ is \coqe{OCtx -> OCtx -> OCtx}.

Wires in a context \coqe{Γ} can be collected into a pattern \coqe{Pat Γ W} to
construct the wire type \code{W}. A pattern is just a tuple of wires of base
types, meaning that all variables in a pattern have type \coqe{Bit} or
\code{Qubit}.  We use Coq's dependent types to express logical
predicates that constrain how patterns can be constructed.
\begin{coq}
Inductive Pat : OCtx -> WType -> Set :=
| unit  : Pat (Valid []) One
| qubit : forall (x : nat) (Γ : Ctx), SingletonCtx x Qubit Γ -> Pat (Valid Γ) Qubit
| bit   : forall (x : nat) (Γ : Ctx), SingletonCtx x Bit   Γ -> Pat (Valid Γ) Bit
| pair  : forall Γ1 Γ2 W1 W2, is_valid(Γ1 ⋓ Γ2)->Pat Γ1 W1->Pat Γ2 W2->Pat (Γ1 ⋓ Γ2) (W1⊗W2).
\end{coq}
The \coqe{pair} constructor (for which we use notation \coqe{(p1,p2)}) ensures
that the wires in \coqe{p1} are disjoint from those in \coqe{p2} by calling out
to the \coqe{is_valid} predicate, which checks whether the result of the merge
is well-defined. The \coqe{qubit} and \coqe{bit} patterns are variable
constructors that are only valid in contexts that contain the exact variable
being introduced. The predicate \coqe{SingletonCtx x W Γ} ensures that the
context $Γ$ contains only the single wire $x$ of type $W$.

\paragraph{Circuits}
\label{sec:flat}

Like patterns, circuits are indexed by an input context and an output type.
There are only three syntactic forms for circuits: \emph{output},
\emph{gate application}, and \emph{dynamic lifting}. \cref{fig:flat} defines
circuits as an inductive data type indexed by the input typing context and the
output wire type.

\begin{figure}
\begin{coq}
Inductive Circuit' : OCtx -> WType -> Set :=
| output' : forall {Γ W}, Pat Γ W -> Circuit' Γ W
| gate'   : forall {Γ Γ1 Γ2 W1 W2 W}, 
            is_valid (Γ1 ⋓ Γ) -> is_valid (Γ2 ⋓ Γ) -> Gate W1 W2 -> 
            Pat Γ1 W1 -> Pat Γ2 W2 -> Circuit' (Γ2 ⋓ Γ) W -> Circuit' (Γ1 ⋓ Γ) W
| lift'   : forall {Γ1 Γ2 W W'}, is_valid (Γ1 ⋓ Γ2) ->
            Pat Γ1 W -> (interpret W -> Circuit' Γ2 W') -> Circuit' (Γ1 ⋓ Γ2) W'.
\end{coq}
\caption{Definition of \qwire circuits using an explicit representation of
  variable binding. We call this type \code{Circuit'}, reserving the name
  \code{Circuit} for the higher-order abstract syntax representation
  (\cref{fig:HOAS}).}
\label{fig:flat}
\end{figure}

An output circuit \coqe{output p} is just a pattern. A gate application, which
we write with syntactic sugar as \coqe{gate_ p2 ← g @p1; C}, is made up of a
gate \coqe{g : Gate W1 W2}, an input pattern \coqe{p1 : Pat Γ1 W1}, an output
pattern \coqe{p2 : Pat Γ2 W2}, and a circuit \coqe{C : Circuit' (Γ1 ⋓ Γ2) W'}.
The intended meaning is that \coqe{p1} is the input to the gate \coqe{g}, and
its output is bound to \coqe{p2} in the continuation \coqe{C}. Thus the
variables \coqe{Γ2} that made up \coqe{p2} are also free in C, whereas the
variables \coqe{Γ1} in \coqe{p1} are no longer available in C, enforcing
linearity. The \coqe{is_valid} predicates enforce that all of the context
arguments to circuits must be well-defined.

The lift operation, which we write \coqe{lift_ x ← p; C} takes as
input a pattern \coqe{p : Pat Γ1 W} and a function \coqe{fun x => C}
that takes the classical interpretation of that data and produces another
circuit. The intended semantics is that the circuit will measure the
wires \coqe{p} (if they are not already bit-valued) and pass the
result as ordinary Coq data to the function. In particular, both bits
and qubits will result in boolean values being provided to the
continuation, and tensors will be interpreted as pairs.

A boxed circuit, written \coqe{box_ p => C} is a pair of a pattern and a circuit.
\begin{coq}
Inductive Box' : WType -> WType -> Set :=
| box' : forall {W1 W2 Γ}, Pat Γ W1 -> Circuit' Γ W2 -> Box' W1 W2.
\end{coq}

\paragraph{Composition and Higher-Order Abstract Syntax}
\label{sec:HOAS}

The minimal embedding of \qwire in \cref{fig:flat} lacks a number of features of
the language, including composition and unboxing. The \coqe{unbox} operator
takes in a boxed circuit and an input pattern and produces a circuit. It has the
following signature:
\begin{coq}
Definition unbox {Γ W1 W2} (b : Box' W1 W2) (p : Pat Γ W1) : Circuit' Γ W2.
\end{coq}
The intended $β$-reduction rule should have the form %
\coqe{unbox (box_ p => C) p' = C[p'/p]} where \coqe{C[p'/p]} is a
substitution of the variables in \coqe{p'} for those in
\coqe{p} in the circuit \coqe{C}.

The traditional solution to this problem involves defining a number of
substitution functions and proving appropriate typing relations for each
operation. Although this approach is viable, it is often tedious and introduces
a large amount of complexity, especially in linear systems. An alternative
approach is the technique of higher-order abstract syntax
(HOAS)~\cite{Pfenning1988}, in which variable bindings in an embedded language
are represented as functions in the host language. This means that the language
designer does not have to define substitution functions and prove their
correctness, and that variables in the embedded language have the same weight as
variables in the host language. 

The HOAS approach to boxed circuits treats a box as a function from patterns to
circuits:
\begin{coq}
Inductive Box : WType -> WType -> Set :=
| box : (forall {Γ}, Pat Γ W1 -> Circuit Γ W2) -> Box W1 W2.
\end{coq}
The \coqe{unbox} operation then simply destructs the box and applies
the function appropriately.
\begin{coq}
Definition unbox {Γ W1 W2} (b : Box W1 W2) (p : Pat Γ W1) : Circuit Γ W2 :=
  match b with box f => f p end.
\end{coq}

\begin{figure}
\begin{coq}
Inductive Circuit : OCtx -> WType -> Set :=
| output : forall {Γ Γ' w}, (Γ = Γ') -> Pat Γ w -> Circuit Γ' w
| gate   : forall {Γ Γ1 Γ1' w1 w2 w}, is_valid Γ1' -> (Γ1' = Γ1 ⋓ Γ)
        -> Gate w1 w2 -> Pat Γ1 w1
        -> (forall {Γ2 Γ2'}, is_valid Γ2' -> (Γ2' = Γ2 ⋓ Γ) -> Pat Γ2 w2 -> Circuit Γ2' w)
        -> Circuit Γ1' w
| lift : forall {Γ1 Γ2 Γ w w'}, is_valid Γ -> (Γ = Γ1 ⋓ Γ2)
        -> Pat Γ1 w -> (interpret w -> Circuit Γ2 w')
        -> Circuit Γ w'.
\end{coq}
\caption{A definition of \qwire circuits using higher-order abstract syntax.}
\label{fig:HOAS}
\end{figure}

We can take a similar approach for the binding pattern in gate application, as
shown in \cref{fig:HOAS}. In this setting, output of a gate application is
represented by a function from the output pattern to a new circuit. The other
difference between the HOAS presentation of circuits and the ``flat''
representation \coqe{Circuit'} in the previous section is that in addition to
proofs of validity about merged contexts, we introduce fresh arguments for the
output context of each circuit. This is due to a technical limitation of Coq
pattern matching, and the result makes it possible for us to write the
examples in \cref{sec:qwire-examples}.

With this machinery in place, we can define composition as a meta-operation on
circuits. For conciseness, we give its definition as a sequence of $β$-reduction
rules, omitting the proof arguments.
\begin{coq}
Fixpoint compose {Γ1 Γ1' W Γ W'}  (c : Circuit Γ1 W) 
         (f : forall {Γ2 Γ2'}, (Γ2' = Γ2 ⋓ Γ) -> is_valid Γ2' -> Pat Γ2 W -> Circuit Γ2' W')
         : is_valid Γ1' -> (Γ1' = Γ1 ⋓ Γ) -> Circuit Γ1' W'.
\end{coq}\begin{coq}
    compose (output p)    f = f p
    compose (gate g p1 h) f = gate g p1 (fun p2 => compose (h p2) f)
    compose (lift p h)    f = lift p (fun x => compose (h x) f)
\end{coq}



%
%

\paragraph{The \qwire type checker}
\label{sec:typechecking}


In type-checking \qwire circuits, we are asked to solve equations of the form
\coqe{Γ1 ⋓ ⋯ ⋓ Γ_n = Γ1' ⋓ ⋯ ⋓ Γ_m'} and \coqe{is_valid (Γ1 ⋓ ⋯ ⋓ Γ_n)}, in
order to enforce the linearity of wires. The first of these goals can be
discharged with a automated proof tactic for solving systems of commutative
monoids. In the higher-order abstract syntax version of circuits, these
predicates may also contain \emph{evars}, existentially quantified Coq
variables. By starting at the leaves of the typing derivation, we can ensure
that every equation of the form \coqe{Γ1 ⋓ ⋯ ⋓ Γ_n = Γ1' ⋓ ⋯ ⋓ Γ_m'} has at most
one evar. We can then cancel out all variables that appear on both sides of the
equation, and unify the evar with what remains on the opposite side.

Once repeated applications of \coqe{monoid} have replaced all existential
variables with regular Coq variables, we can prove goals of the form
\coqe{is_valid (Γ1 ⋓ ⋯ ⋓ Γ_n)}. We take advantage of the fact that a set of
finite contexts is disjoint if and only if the contexts are all pairwise
disjoint. Our custom tactic for solving these goals extracts all proofs of
pairwise disjointness from hypotheses and then reduces the goal to a conjunction
of pairwise disjointness claims \coqe{is_valid(Γ_i ⋓ Γ_j)}. It then applies
proofs of these claim, where available, from the hypotheses.

\section{Denotational Semantics}
\label{sec:denotation}
\subsection{The Matrix Library}

The denotational semantics of \qwire is implemented using a matrix library
created specifically for this purpose. Matrices are simply functions from pairs
of natural numbers to complex numbers.\footnote{As a Coq technicality, note that
  matrices are only equal up to functional extensionality.}
\begin{coq}
Definition Matrix (m n : nat) := nat -> nat -> \CC.
\end{coq}
The arguments $m$ and $n$, which are the dimensions of the matrix, are not used
directly in the definition, but they are useful to define certain operations on
matrices, such as the Kronecker product and matrix multiplication, which depend
on these dimensions. They are also useful as an informal annotation that aids
the programmer. We say a matrix is well-formed when it is zero-valued outside of
its domain.
\begin{coq}
Definition WF_Matrix {m n} (M : Matrix m n) : Prop := forall i j, i >= m \/ i >= n -> M i j = 0.
\end{coq}

This library is designed to facilitate reasoning about and computing with
matrices. Treating matrices as
functions allows us to easily express otherwise complicated matrix operations.
Consider the definitions of Kronecker product (⊗) and complex
conjugate transpose (†), where \coqe{Cconj} is the complex conjugate:
\begin{coq}
Definition kron {m n o p} (A : Matrix m n) (B : Matrix o p) : Matrix (m*n) (o*p) :=
  fun x y => A (x / o) (y / p) * B (x mod o) (y mod p).
Definition ctrans {m n} (A : Matrix m n) : Matrix n m := fun x y => Cconj (A y x).
\end{coq}
We represent complex numbers using an adaptation of Coquelicot's \coqe{Complex} library~\cite{coquelicot}. 
This representation has the advantage of
being a straightforward extension of the Coq real numbers, allowing us to easily
extend the Coq Standard Library's powerful Linear Real Arithmetic solver
(\texttt{lra}) to both complex numbers and matrices. The tactics we define
(termed \texttt{clra} for complex numbers and \texttt{mlra} for matrices) allow
us to trivially prove that complex conjugate and complex conjugate transpose are
involutive:
\begin{coq}
Lemma conj_involutive : forall (c : \CC), Cconj (Cconj c) = c.  Proof. intros. clra. Qed.
Lemma ctrans_involutive : forall {m n} (A : Matrix m n), A†† = A. Proof. intros. mlra. Qed.  
\end{coq}



\subsection{Density Matrices}

\qwire programs are interpreted as superoperators over density matrices,
following the denotational semantics described by
Paykin~\etal~\cite{Paykin2017}. We use a density matrix representation over the
standard unit vector representation (used for example by
Boender~\etal~\cite{Boender2015}) because density matrices represent probability
distributions (introduced via measurement) over quantum states directly, as
opposed to the unit vector representation which must be embedded inside a
probability monad.

We start with some preliminary definitions. A unitary matrix is a well-formed
square matrix $A$ such that $A^†×A$ is the identity.
\begin{coq}
Definition is_unitary {n} (A : Matrix n n) := WF_Matrix A  /\  A† × A = Id n.
\end{coq}

A pure state of a quantum system is one that corresponds to a unit vector
$\ket{φ}$. An equivalent representation is that of square matrices $ρ$ such that
$ρ × ρ = ρ$.
\begin{coq}
Definition Pure_State {n} (\rho : Matrix n n) : Prop := WF_Matrix ρ  /\  ρ = ρ × ρ.
\end{coq} 

A density matrix, or mixed state, is a linear combination of pure states
representing the probability of each pure state.
\begin{coq}
Inductive  Mixed_State {n} (ρ : Matrix n n) : Prop :=
| Pure_S : forall ρ, Pure_State ρ -> Mixed_State ρ
| Mix_S  : forall (p : \RR) ρ1 ρ2, 0 < p < 1 
        -> Mixed_State ρ1 -> Mixed_State ρ2 -> Mixed_State (p .* ρ1 .+ (1-p) .* \rho2).
\end{coq}
Note that every mixed state is also well-formed, since scaling and addition preserve well-formedness.

A superoperator is a function on square matrices that takes mixed states to
mixed states.
\begin{coq}
Definition Superoperator m n := Matrix m m -> Matrix n n.
Definition WF_Superoperator m n (f : Superoperator m n) := 
    forall (ρ : Matrix m m), Mixed_State ρ -> Mixed_State (f ρ).
\end{coq}
Any $m × n$ matrix can be lifted to a superoperator from $n$ to $m$ as follows:
\begin{coq}
Definition super {m n} (A : Matrix m n) : Superoperator n m := fun ρ => A × ρ × A†.
\end{coq}

\subsection{Denotation of \qwire}

 \paragraph{Types, Contexts and Gates}

In order to interpret circuits as superoperators over density matrices, we will
also give types, contexts, gates, and patterns interpretations in linear
algebra. For clarity we write \coqe{〚-〛} for the denotation of a variety of
\qwire objects, which we express via a Coq type class.
\begin{coq}
Class Denote source target := { denote : source -> target }.
Notation "[[ x ]]" := (denote x) (at level 10).
\end{coq}

We interpret every wire type as the number of primitive (\coqe{Bit} or
\coqe{Qubit}) wires in that type, so \coqe{〚Qubit ⊗ (One ⊗ Bit)〛 = 2}.
Contexts and \coqe{OCtx}s are similarly denoted by the number of wires they
contain.

Every gate of type \coqe{Unitary W} corresponds to a unitary matrix of
dimension $2^{〚W〛} × 2^{〚W〛}$. We omit the implemented gate set and their
corresponding matrices here, but in the development we prove that every denoted
unitary satisfies the \coqe{is_unitary} predicate defined above.
\begin{coq}
Lemma unitary_gate_unitary : forall {W} (u : Unitary W), is_unitary [[u]].
\end{coq}

A gate of type \code{Gate W1 W2} corresponds to a superoperator as follows:
\begin{coq}
Definition denote_gate {W1 W2} (g : Gate W1 W2) : Superoperator 2^[[W1]] 2^[[W2]] :=
  match g with
  | U u          => super [[u]]
  | init0 | new0 => super |0⟩
  | init1 | new1 => super |1⟩
  | meas         => fun ρ => super |0⟩⟨0| ρ .+ super |1⟩⟨1| ρ
  | discard      => fun ρ => super ⟨0| ρ .+ super ⟨1| ρ
  end.
Instance Denote_Gate {W1 W2} : Denote (Gate W1 W2) (Superoperator 2^[[W1]] 2^[[W2]]) :=
   {| denote := denote_gate |}.
\end{coq}

When applying a gate to a subset of a quantum system, however, we will need a
generalization of the \coqe{denote_gate} operation that applies the gate to
the first part of a quantum system.
\begin{coq}
Definition denote_gate' n {W1 W2} (g : Gate W1 W2) : Superoperator 2^[[W1]]*2^n 2^[[W2]]*2^n.
\end{coq}
What we previously wrote as \coqe{denote_gate} is simply \coqe{denote_gate' 0}.



\paragraph{Patterns and Flat Circuits}

\begin{wrapfigure}{r}{0.2\textwidth}
\[\code{swap} = \begin{pmatrix} 
1 & 0 & 0 & 0 \\
0 & 0 & 1 & 0 \\
0 & 1 & 0 & 0 \\
0 & 0 & 0 & 1 
\end{pmatrix} \]
\end{wrapfigure}

Patterns of type \coqe{Pat Γ W} are interpreted as permutation matrices of
dimension $2^{〚Γ〛} × 2^{〚W〛}$. These matrices are constructed via multiple
applications of a \coqe{swap} matrix, as follows: 
In general we will want to swap the positions of two arbitrary qubits in a
system; to swap qubit 0 with qubit 2 in a 3-qubit system, we invoke
$\code{swap2 0 2} = (I_2 ⊗ \code{swap})(\code{swap} ⊗ I_2)(I_2 ⊗ \code{swap})$. 

First, every pattern is interpreted as a list of indices indicating the wire
number (counting from $0$) that each variable refers to. So the pattern
\coqe{(3,0) : Pat [Some Qubit, None, None, Some Bit] (Bit ⊗ Qubit)} corresponds to the
list $[1;0]$, because variable $3$ corresponds the wire numbered $1$ and variable
$0$ corresponds to wire $0$ in the circuit. This list \coqe{ls} is then turned into
an association list $[(0,1);(1,0)]$, mapping variable \coqe{i} to \coqe{ls[i]}. These pairs
are then interpreted as a series of calls to \coqe{swap2}.

As for gates, the function \coqe{denote_pat_in} follows a similar algorithm, but
allows us to interpret a pattern \coqe{p} inside a larger context of variables.
Its signature is as follows:
\begin{coq}
Definition denote_pat_in Γ' {Γ W} (p : Pat Γ W): Matrix 2^〚Γ ⋓ Γ'〛 (2^〚W〛 * 2^〚Γ'〛).
Instance Denote_Pat {Γ W} : Denote (Pat Γ W) (Matrix 2^〚Γ〛 2^〚W〛) := 
    {| denote := denote_pat_in (Valid []) |}.
\end{coq}


Finally, we interpret circuits as superoperators on density matrices. To do
this, we start with the ``flat'' representation \coqe{Circuit'} that does not
use higher-order abstract syntax. The higher-order abstract syntax
representation is useful for defining the meta-level operations on circuits, but
the ``flat'' circuits from \cref{fig:flat} use concrete variables, thus making
them a useful intermediate representation for the semantics.

Consider an output circuit \coqe{output' p}, where \coqe{p : Pat Γ W}. The
interpretation of this circuit is the superoperator obtained from the denotation
of \coqe{p}:
\coqe{〚output' p〛 = super 〚p〛.}

Next, consider \coqe{gate_ p2 <- g @p1; C}, where %
\coqe{g : Gate W1 W2}, 
\coqe{p_i : Pat Γ_i W_i} and 
\coqe{C : Circuit' (Γ2 ⋓ Γ) W}. The interpretation of \coqe{C} has type
\coqe{Superoperator 2^〚Γ2 ⋓ Γ〛 2^〚W〛}, so we need to compose \coqe{〚C〛}
with a superoperator from 
\coqe{2^〚Γ1〛*2^〚Γ〛} to \coqe{2^〚Γ2〛*2^〚Γ〛}
which
we obtain by composing \coqe{denote_gate'} with the results of
\coqe{denote_pat_in}, to rearrange the quantum system appropriately:
\begin{coq}
    〚gate' g p1 p2 C'〛 = 〚C'〛 ∘ super (〚p2〛† ⊗ Id 2^〚Γ〛) 
                             ∘ denote_gate' 〚Γ〛 g ∘ super (〚p1〛 ⊗ Id 2^〚Γ〛)
\end{coq}

Finally, consider a lift circuit, \code{lift' p f}, where \coqe{p : Pat Γ1 W}
and \coqe{f : interpret W -> Circuit' Γ2 W'}. When \coqe{W} is a qubit,
\coqe{interpret W = bool}, and the lift operation would measure the qubit and
sum over the results of \coqe{〚f true〛} and \coqe{〚f false〛}. More
generally, consider %
\begin{coq}
Definition f' : interpret W -> Superoperator 2^〚Γ1 ⋓ Γ2〛 2^〚Γ2〛
    := fun x => 〚f x〛 ∘ (super ((kets x)† ⊗ Id 2^〚Γ2〛)) ∘ (super (denote_pat_in Γ2 p))
\end{coq}
Here, \coqe|kets {W} : interpret W -> Matrix 2^〚W〛 1| is the basis
representation of the input value of type \coqe{interpret W}. By transposing
\coqe{kets x} and expanding it via the \coqe{super} operation, we pick out the
partial density matrix corresponding to that measurement branch. Next, since all
wire types are finite, we can enumerate all values of type \coqe{interpret W} in
a list via the operation \coqe{get_interpretations W}. 
By mapping \coqe{f'} over this list, we obtain each of the actual measurement
branches as superoperators. Now we can simply perform pointwise addition of the
superoperators, and compose with the pattern \coqe{p} to organize the wires in
order:
\begin{coq}
    〚lift' p f〛 = fold_left Splus (map f' (get_interpretations W)) SZero
\end{coq}

A flat box is also interpreted as a superoperator:
\coqe{ 〚box' p C〛 = 〚C〛 ∘ super 〚p〛†}

\paragraph{HOAS Circuits}

To denote the HOAS version of circuits, we first map them to our
representation of ``flat'' circuits, which involves instantiating the
output patterns of HOAS gates with a particular concrete pattern. We
do this via an operation \code{fresh\_pat} that takes as input a
context and a type, and produces a pattern of that type whose domain
(\coqe{fresh_pat_ctx}) is disjoint from the input context.
\begin{coq}
Definition fresh_pat (Γ : OCtx) (W : WType) : Pat (fresh_pat_ctx Γ W) W.
\end{coq}

Using \coqe{fresh_pat} we can define a function that converts HOAS circuits and
boxes into flat circuits and boxes. We leave off implicit proof and \coqe{OCtx}
arguments for legibility.
\begin{coq}
Program Fixpoint from_HOAS {Γ W} (c : Circuit Γ W) : Circuit' Γ W :=
 match c with
 | output p    => output' p
 | gate g p1 f => gate' g p1 p2 (from_HOAS (f (fresh_pat _ _)))
 | lift p f    => lift' p (fun x => from_HOAS (f x))
 end.
Program Definition from_HOAS_Box {W1 W2} (b : Box W1 W2) : Flat_Box W1 W2 :=
  match b with box f => let p := fresh_pat [] W1 in box' p (from_HOAS (f p)) end.
\end{coq}

\noindent The denotation of a HOAS circuit is exactly the denotation of its
corresponding flat circuit.


\section{A Taste of Verification}
\label{sec:verification}




When a circuit is closed, that is when it has no input, it represents a
preparation of a quantum state. In many cases, a programmer may know what state
their program should prepare, and our verification framework allows them to
compare the denotation of the circuit with the desired density matrix directly. 

Consider, for instance, the coin flip circuit in \cref{sec:qwire-examples}. In
the online development we prove that the denotation of \coqe{coin_flip}
corresponds to the matrix
\coqe{even_toss}$=\left(\begin{smallmatrix}\rfrac{1}{2} & 0 \\ 0 & \rfrac{1}{2}\end{smallmatrix}\right)$ as
  follows:
\begin{coq}
Lemma flip_toss : [[ coin_flip ]] (Id 1)  = even_toss.
\end{coq}
The Coq proof script used to prove this lemma is only twelve lines long, and
calls out to a number of specialized tactics, including \texttt{Msimpl},
\texttt{Csimpl} and \texttt{Csolve}, designed to simplify and prove equality
between matrices and complex numbers. 

The combination of Coq and \qwire truly shines in its ability to prove more
complex properties and prove properties of \emph{families} of circuits, not
merely circuits themselves.
For example, consider the Deutsch-Jozsa problem, where we want to verify a
property of a family of circuits with any number of input qubits, that holds for
every appropriate unitary matrix $U_f$. This sort of property requires induction
over the number of inputs and the full power of a proof assistant.


Open circuits, which correspond to arbitrary superoperators, are even more
interesting from the perspective of verification. Consider the following circuit
which composes a unitary gate with its transpose:

\begin{minipage}{0.65\textwidth}
\begin{coq}
Definition unitary_trans {W} (U : Unitary W) : Box W W.
  box_ p =>
    gate_ p ← U @p;
    gate_ p ← transpose U @p;
    output p.
Lemma unitary_trans_id : forall W (U : Unitary W) ρ,
    WF_Matrix (2^〚W〛) (2^〚W〛) ρ -> 〚unitary_trans U〛ρ = 〚id_circ W〛ρ.
\end{coq}
\end{minipage}
\begin{minipage}{0.3\textwidth}
\[
\Qcircuit @C=2em @R=-1em {
& \multigate{4}{\mathcal{U}} & \multigate{4}{\mathcal{U^\dag}} & \qw \\
& \ghost{\mathcal{U}} & \ghost{\mathcal{U^\dag}} & \qw \\
& \ghost{\mathcal{U}} & \ghost{\mathcal{U^\dag}} & \qw \\
& \ghost{\mathcal{U}} & \ghost{\mathcal{U^\dag}} & \qw \\
& \ghost{\mathcal{U}} & \ghost{\mathcal{U^\dag}} & \qw
} \]
\end{minipage}

\noindent
This correctness of this circuit holds for any unitary gate regardless of its
input size. The proof of this lemma takes fewer than $20$ lines of code, and
proceeds using common facts from linear algebra, such as the fact that %
\coqe{Id n × A = A}, along with fact that the denotation of \coqe{U} is in fact
unitary, that is, \coqe{〚U〛† 〚U〛 = Id}. We also rely on the fact that the
denotation of every pattern in this circuit is actually an identity matrix.

\begin{wrapfigure}{r}{0.52\textwidth}
\vspace{-1em}
\begin{coq}
Definition lift\_meas : Box Qubit Bit.
  box_ q =>
    lift_ x ← q;
    gate_ p ← (if x then new1 else new0) @();
    output p.
Lemma lift\_meas_correct : forall ρ, WF_Matrix ρ 
   -> 〚lift\_meas〛 ρ = 〚boxed_gate meas〛 ρ.
\end{coq}
\end{wrapfigure}
Another useful sanity check ensures that the denotation of a \coqe{lift} circuit
is equivalent to applying a measurement gate. For example, consider
\coqe{lift\_meas}, which measures a qubit via \coqe{lift} and then reconstructs
a bit-valued wire from the result of the measurement.
The resulting circuit is equivalent to the circuit that simply applies a
measurement gate.



\section{Related and Future Work}
\label{sec:conclusion}

The Coq development described in this paper is very much still a
work in progress. There are a small number of lemmas in the underlying matrix
library that we have not yet formally proved, although they are known facts about
linear algebra. In addition, we have not yet formally proved that the denotation
of every circuit is a well-formed superoperator over density matrices, or that
the denotation of the composition of two circuits is equal to the composition of
their denotations. 

After completing this work, there are a number of exciting areas to explore in
the future.

\paragraph{Verified Compilation}

By verifying the equivalence of circuit transformations like
\coqe{unitary_trans_id} from \cref{sec:verification}, we see \qwire as a
prime language in which to compile quantum programs. The area of verified
compilers has gained a lot of traction in recent years, inspired by the success
of the CompCert C compiler~\cite{Leroy2004}. Towards this 
direction, Amy~\etal~\cite{Amy2017} developed
a verified, lightly-optimizing compiler for reversible circuits, which can
encompass unitary quantum circuits. 

We can also provide insight into a relationship between \qwire circuits and
QASM~\cite{QASM}, a quantum assembly-like language that has gained widespread
use in quantum simulators and IBM's Quantum Experience~\cite{IBM}. The largest
difference between \qwire and QASM is our use of variable binding, using
abstract variables such as $x$, $y$, and $z$ and allowing qubits to be renamed
as in %
\coqe{let_ y ← output x; C'}. In comparison, QASM operates on a concrete set of
quantum registers (\eg qubits 1, 2, or 3) that cannot be renamed. In the
development we provide a version of \qwire that similarly operates directly on
named qubits. Our denotational semantics extends directly to this
representation, and we can compile from \qwire to these ``assembly-level''
circuits. Future work could formally establish a relationship between the
assembly-level version of \qwire and QASM, allowing us to prove properties of
QASM programs, and prove that compilation is sound with respect to the
denotational semantics.

\paragraph{A More Efficient Backend}

Unfortunately, we have so far struggled to prove properties of
(non-parametric) circuits with more than a few qubits, indicating that
scalability will be a challenge moving forward. A key contributor to the
scalability of our theorems is the underlying matrix library, which is not
currently optimized in any significant way. Of course, any quantum simulator
will be intractable on large enough circuits, as density matrices are always
exponential in the size of the corresponding circuit.

In the near future, we will be transitioning the linear algebra back-end of our
development to one of several existing projects, in the hopes of making it more
efficient. Boender~\etal~\cite{Boender2015} present one candidate, a library
developed reasoning about the correctness of quantum protocols using pure
states. Another candidate is the Coq Effective Algebra Library
(CoqEAL)~\cite{coqeal}, which is designed specifically to allow matrix
computation inside Coq. This library allows easy translation between it matrices
and those of the Mathematical Components library~\cite{mathcomp}, which in turn
was designed with an eye towards verification. The two libraries together may
substantially increase our ability to run and reason about \qwire programs. We
can also simulate \qwire programs by extraction to OCaml.

\paragraph{Theory of \qwire}

In this paper we focus on the denotational semantics of \qwire, but many other
aspects of the meta-theory are left to be explored. Rennela and
Staton~\cite{Rennela2017} present a categorical model of EWire, a close variant of
\qwire, as an enriched category. Their model also allows for additional features
such as quantum data types in the style of Quipper~\cite{Green2013}. 

The equational theory of quantum circuits is another area left for future work.
For example, Staton~\cite{Staton2015} presents an axiomatization of the
relationship between measurement, qubit initialization, and a limited set of
unitary gates. In the future we hope to adapt Staton's work to \qwire and
thereby reason syntactically, rather than semantically, about the equivalence of
\qwire circuits. An equational theory is also key to integrating \qwire with
dependent types~\cite{Paykin2017}.

\paragraph{Verifying Higher-Order Programs}

The use of dependent types was a driving factor for both the development of
\qwire and the choice to embed it in Coq specifically. While dependent types
power Coq's verification capabilities, they're also key to our representation of
circuits. For example, in the development we implement a dependently-typed 
version of the Quantum Fourier Transform, as described in the introduction to 
\qwire~\cite{Paykin2017}. 
This paper hasn't focused on dependent types, or high level programming
in \qwire generally, but both will feature heavily in future \qwire development.

Greater use of the host language and classical programming abstractions will
also allow to further push the boundaries of quantum verification. While current
\qwire programs can be thought of as ``simply circuits,'' such direct
proofs will become increasingly difficult as we add on layers of abstraction---just as it would be
nearly impossible to prove interesting program properties by reasoning about the
underlying classical circuits. This will require new approaches to quantum
verification, from quantum weakest precondition reasoning~\citep{Dhondt2006} 
(expanded into a Hoare-like logic in \citep{Ying2011, Ying2017}), to the forms of
equational reasoning described above.




\bibliographystyle{eptcsstyle/eptcs}
{\small
\bibliography{biblio}
}

\end{document}